\documentclass{kluwer}    
\usepackage{epsfig}

\newdisplay{guess}{Conjecture}

\begin{document}              
                                                                     
\begin{article}
\begin{opening}         
\title{Status and First Results of the MAGIC Telescope} 
\author{Juan \surname{Cortina} for the MAGIC collaboration}  
\runningauthor{Juan Cortina}
\runningtitle{Status of MAGIC}
\institute{Institut de Fisica d'Altes Energies, Universitat
Aut\`{o}noma de Barcelona, Bellaterra 08190, Spain}
\date{May 26, 2004}

\begin{abstract}
The 17 m MAGIC Cherenkov telescope for gamma ray astronomy between
30 and 300 GeV started operations in its final configuration in
October 2003 and is currently well into its calibration phase.
Here I report on its present status and its first gamma ray 
source detections.
\end{abstract}

\end{opening}           

\section{Introduction}
The MAGIC (Major Atmospheric Gamma Imaging Cherenkov) Telescope
was designed in 1998 \cite{Barrio} with the main goal of being the Imaging
Atmospheric Cherenkov Telescope (IACT) with the lowest possible gamma
energy threshold. It was based on the experience
acquired with the first generation of Cherenkov telescopes
and intended to incorporate a large number of technological improvements.

There was a clear-case motivation to head for the low energy
threshold: there was a well populated sky-map of sources 
detected by EGRET (but more than
half of them still unidentified due to poor angular resolution)
around 10 GeV in contrast to the a handful of sources observed by
the existing IACTs above 300 GeV. The main idea was to cover the
unexplored energy gap in between with an IACT.
These detectors provide much large effective areas 
(around 4$\cdot$10$^4$ m$^2$) than satellite detectors, better angular resolution
(ranging from 0.2$^{\circ}$ close to the threshold down to
0.2$^{\circ}$ at higher energies), acceptable energy
resolution (going down from 30\% at the threshold energy to
less than 20\% above 100 GeV) and a well tested capability to separate 
gammas from other backgrounds.

From the purely experimental point of view, covering that gap with IACTs could allow:

\begin{itemize}
\item{} \vspace{-0.25 cm} To study the mechanisms which cut-off
the spectrum of several of the EGRET sources precisely in this
energy gap and to explain why they were not detected by the
first generation of IACTs above 300 GeV.

\item{} \vspace{-0.25 cm} To study all the EGRET sources with a
much higher flux sensitivity and angular resolution and hence,
identify the EGRET unidentified sources.

\item{} \vspace{-0.25 cm} To eventually discover a plethora of new
sources \cite{Barrio} since for most of known sources the energy spectrum is of
power-law nature and therefore they should exhibit a much higher
flux at lower energies. 
\end{itemize}

Several innovative technical solutions started being worked out as
early as in \cite{Lorenz} and, since then the R{\&}D has not stopped.
During this time several options were discussed, such as the
convenience of a single very large IACT incorporating the latest
technological developments or a solution based on an array of
somewhat smaller conventional telescopes. A cost and physics
comparison led to the conclusion that a single very large diameter
IACT would be cheaper and would allow us to cope better with the
prime goal of reducing the threshold as much as possible. We were
nevertheless aware that the single large telescope choice had some
drawbacks: a less efficient background rejection, a
somewhat worse sensitivity at higher energies and a
somewhat poorer angular and energy resolution than a system of
telescopes. 

Nevertheless we deem that these drawbacks are offset
by the exploratory character of our telescope. From an instrumental
point of view this is a telescope whose main intention is 
to bring the Imagining Cherenkov Technique below 100~GeV 
and from the astrophysical point of view it intends to go as deep
as possible into the unexplored gap and get as close in energy
as possible to the range that was explored by EGRET.

\section{Description and Status of the Telescope}

MAGIC is a large and light-weight Cherenkov telescope 
which incorporates a large number of technical innovations. 
It is located at the
Roque de los Muchachos Observatory (ORM) at 2200 m asl (28.8$^o$ north,
17.9$^o$ west) on the Canary island of La Palma.

\subsection{The frame and the drive system}

The 17~m diameter $f/D = 1$ telescope frame is made by light weight
carbon fiber tubes (the frame itself weighs $<$~20~ton while the
whole structure plus the undercarriage amounts to about 60~ton). 
The construction of the foundation for the MAGIC telescope started
in September 2001 and just a few months later the whole telescope
structure was completed (December 2001).
In fact, the assembly of the whole frame took only one month because 
of a construction based on the so-called tube and knot system of the company MERO.

The telescope drive system of the alt-az was installed during 2002. 
The azimuth axis is equipped with two 11~kW motors,
while the elevation axis has a single motor of the same power.
The position of the telescope is measured in the mechanical telescope
frame by three absolute 14-bit shaft encoders. With this configuration 
it is possible to measure the telescope position with an accuracy of about
0.02$^{\circ}$. The maximum repositioning time of the telescope is 22
seconds, well below the 30 seconds target required for gamma ray burst 
follow-up. By using a CCD camera mounted on the reflector frame we have
established that the telescope tracks to better than a 1/10 of a pixel size.


\subsection{The Reflector}

The overall reflector shape is parabolic to minimize the 
time spread of the Cherenkov light flashes in the camera plane.
The preservation of the time structure of the Cherenkov pulses
is important to increase the signal to noise ratio with respect
to the night-sky background light (NSB). The dish is tessellated by 
956 0.5~$\times$~0.5~m$^2$ mirrors covering a total surface of 234~m$^2$.
Each mirror is a sandwich of aluminium honeycomb on which a 5~mmm
plate of AlMgSi1.0 alloy is glued. The aluminium plate is 
diamond-milled to achieve a spherical reflecting surface with the
radius of curvature that is more adequate for its position in the
paraboloid. A thin quartz layer protects the mirror surface from aging.
The reflectivy of the mirrors is around 90\%.
The alignment of the mirrors in the telescope surface has been done
using an artificial light source at a distance of 960~m. The overall
spot has a FWHM of roughly half a pixel size ($<$~0.05$^{\circ}$).

\subsection{The Active Mirror Control}

A large diameter telescope has strong requirements on the stiffness
of the reflector frame. When directing the telescope to different 
elevation angles the reflector's surface deviates from its ideal shape
under gravitational load. Two solutions are possible to counteract
the deformations: a) either to construct a very heavy and stiff frame
or b) allow for small deformations by constructing a light-weight
structure and correct its mirror profile. We have chose the second option for
the MAGIC telescope and equipped the reflector with an ``Active Mirror
Control'' system. 
Each four mirror facettes are mounted on a single panel.
Two of the three mounting points of the panel are equipped with actuators 
which can be used to adjust its position on the frame. 
The main elements of each actuator are
a two-phase stepping motor (full step 1.8$^{\circ}$, holding torque 50~N~cm)
and a ballspindle (pitch 2~mm, maximum range 37~mm). In the center
of the panel a laser module is pointed towards the common focus of the four
mirrors. The panels are aligned using the artificial light source,
the positions of all the laser spots are recorded and can be used as a reference
to re-align them for each elevation angle. 
The AMC has undergone extensive tests during winter 2003/04
and it is presently close to its nominal performance.
The light spot of bright stars focussed at infinity fit
within one pixel (RMS of the Point Spread Function 
less than 0.1$^{\circ}$).

\subsection{The Camera}

The MAGIC camera has 1.5~m diameter, 450~kg weight and 3-4$^{\circ}$~FOV.
The inner hexagonal area is composed by 397~0.1$^{\circ}$ FOV hemispherical 
photomultipliers of 1~inch diameter (Electron Tubes 9116A\cite{Ostankov}) 
surrounded by 180 0.2$^{\circ}$ FOV PMTs of 1.5~inch diameter (ET 9116B).
The time response FWHM is below 1~ns. The photocathode quantum efficiency
is enhanced up to 30\% and extended to the UV by a special coating of the
surface using wavelength shifter \cite{Paneque}. 
Each PMT is connected to an
ultrafast low-noise transimpedance pre-amplifier, the 6-dynode high voltage
system is stabilized with an active load. Dedicated light collectors have been
designed to let the photon double-cross the PMT photocathode for large 
acceptance angles. The anode current and HV are read out for each pixel
and digitized by a 12 bit ADC. The temperature and humidity are controlled
by a water based cooling system. The camera was completed in summer 2002
after extensive testing and characterization; it was installed on
the site in November 2002 and commissioned in March 2003.
First starlight using the DC current
readout was recorded already on 8th of March 2003. In the last
year the camera has complied with its expected performance.

%
%
%

\subsection{The Readout System}

The PMT signals are amplified at the camera and transmitted over 162~m long
optical fibers using Vertical Cavity Emitting Laser Drivers (VCSELs, 
850~nm wavelenght). Transmission over optical links drastically reduces
the weight and size of the cables and protects the Cherenkov signal 
from ambient electromagnetic noise in the line. In the receiver boards in
the electronics room the signal is amplified and split. One branch goes to a
software adjstable threshold discriminator that generates a digital signal
for the trigger system. The signal in the second branch is 
stretched to 6~ns FWHM and again split into a high gain line where it is
further amplified by a factor $\sim$10 while the low gain line is just
delayed by 50~ns. If the signal is above a preset threshold both lines
are combined using a GaAs analog switch and digitized by the same FADC channel.

The 8~bit 300~MHz Flash ADCs continuously digitize the analog signals
and store the digital data into a 32~kByte long ring buffer. If a
trigger signal arrives within less than 100~$\mu$s the position
of the signal in the ring buffer for each pixel is determined  and for each
pixel 15~high gain plus 15~low gain samples are written to a 512~kByte
long FIFO buffer at a maximum rate of 80~Mbyte/s. The readout of the ring
buffer results in a dead time $\sim$20~$\mu$s. This corresponds 
to about 2\% dead time at the design trigger rate of 1~kHz.
The time and trigger information for each event are recorded by dedicated 
digital modules which are read out along with the FADC analog modules.
The readout is controlled by an FPGA (Xilinx) chip on a PCI (MicroEnable)
card. The data are saved to a RAID0 disk system at a rate up to 20~MByte/s
which results in up to 800 GByte raw data per night. 

\subsection{The Trigger}

MAGIC is equipped with a two-level trigger system with programmable logic 
\cite{Bastieri}. The first level (L1T) applies tight time coincidence
and simple next neighbour logic. The trigger is active in 19 hexagonal
overlapping regions of 36 pixels each, to cover 325 of the inner pixels
of the camera. The second level (L2T) can be used to perform a rough analysis
and apply topological constraints on the event images. Using for instance
a fast evaluation of the size of the Cherenkov image it is possible to 
reduce significantly the NSB, thus allowing a reduction
of the discrimination level and the gamma ray threshold. The individual
pixel rates of the channels included in the trigger are monitored
using 100~MHz scalers.

The 1st and 2nd level trigger systems were installed
and commissioned during 2003 as well as the whole computing system for the
telescope control and DAQ. 

\subsection{The calibration system}

The calibration system consists of a light pulser and a continuous light
source (both situated in the center of the mirror dish), a darkened,
single photoelectron counting PMT (``blind pixel''), located in the
camera plane and a calibrated PIN-diode 1.5 meter above the light pulser.
The pulsed light is emitted by very fast (3-4~ns FWHM) and powerful
(10$^8$-10$^{10}$~photons/sr) light emitting diodes in three different
wavelenghts (370~nm, 460~nm and 520~nm) and different intensities
(up to 2000-3000 photoelectrons per pixel and pulse). It is threfore possible to
calibrate the whole readout chain in wavelength and linearity. The
continuous light source consists of LEDs simulating the NSB
at La Palma with different intensities. The light pulser and
continuous light source were commissioned during the winter 2003/04,
while the blind pixel and PIN diode are currently in their commissioning
phase.

\section{First Data}

\subsection{Performance}

\begin{figure}
\begin{minipage}[b]{.45\linewidth}
  \begin{center}
    \includegraphics[height=12pc]{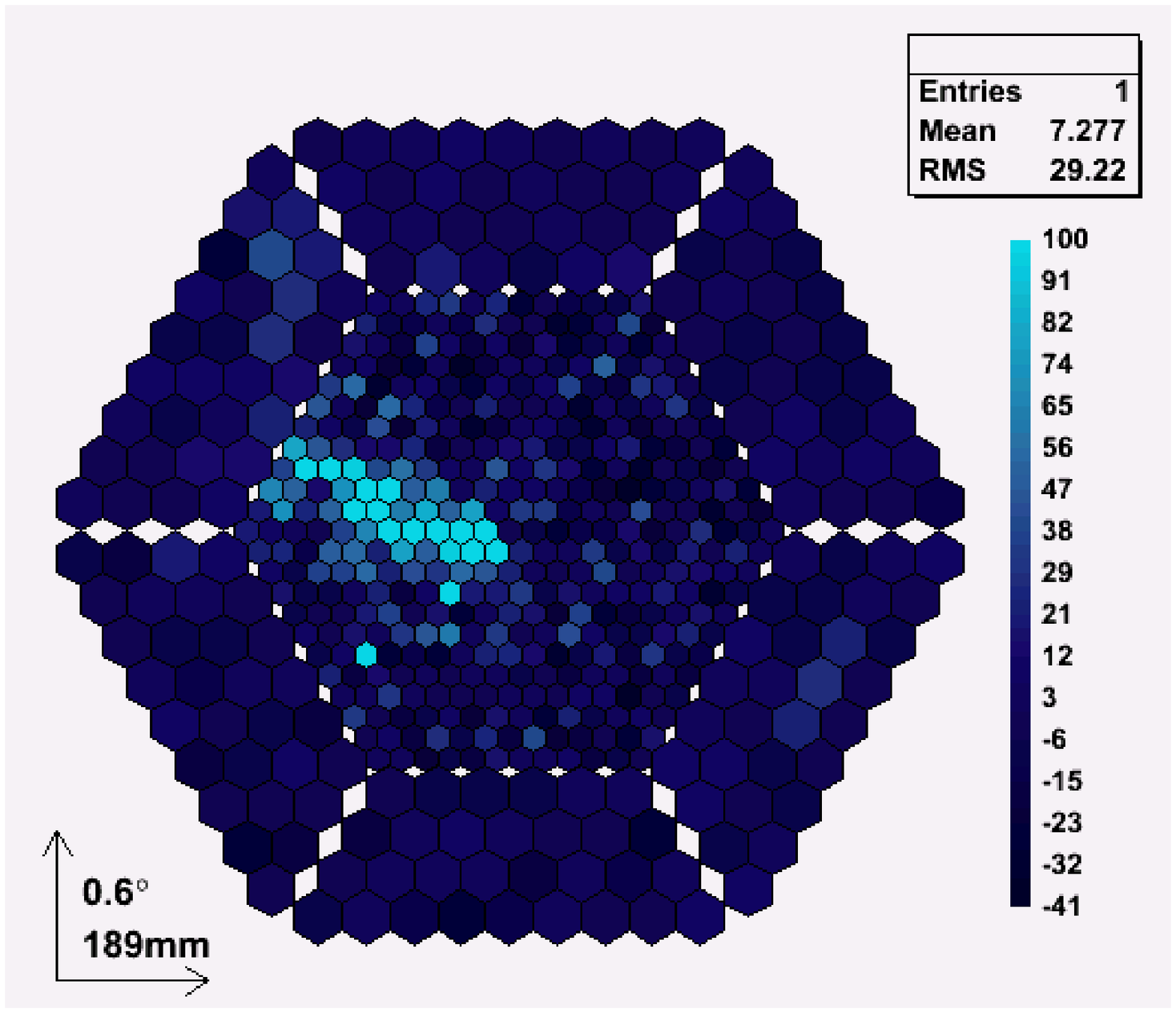}
  \end{center}
  \vspace{-0.5pc}
\end{minipage}\hfill
\begin{minipage}[b]{.45\linewidth}
  \begin{center}
    \includegraphics[height=12pc]{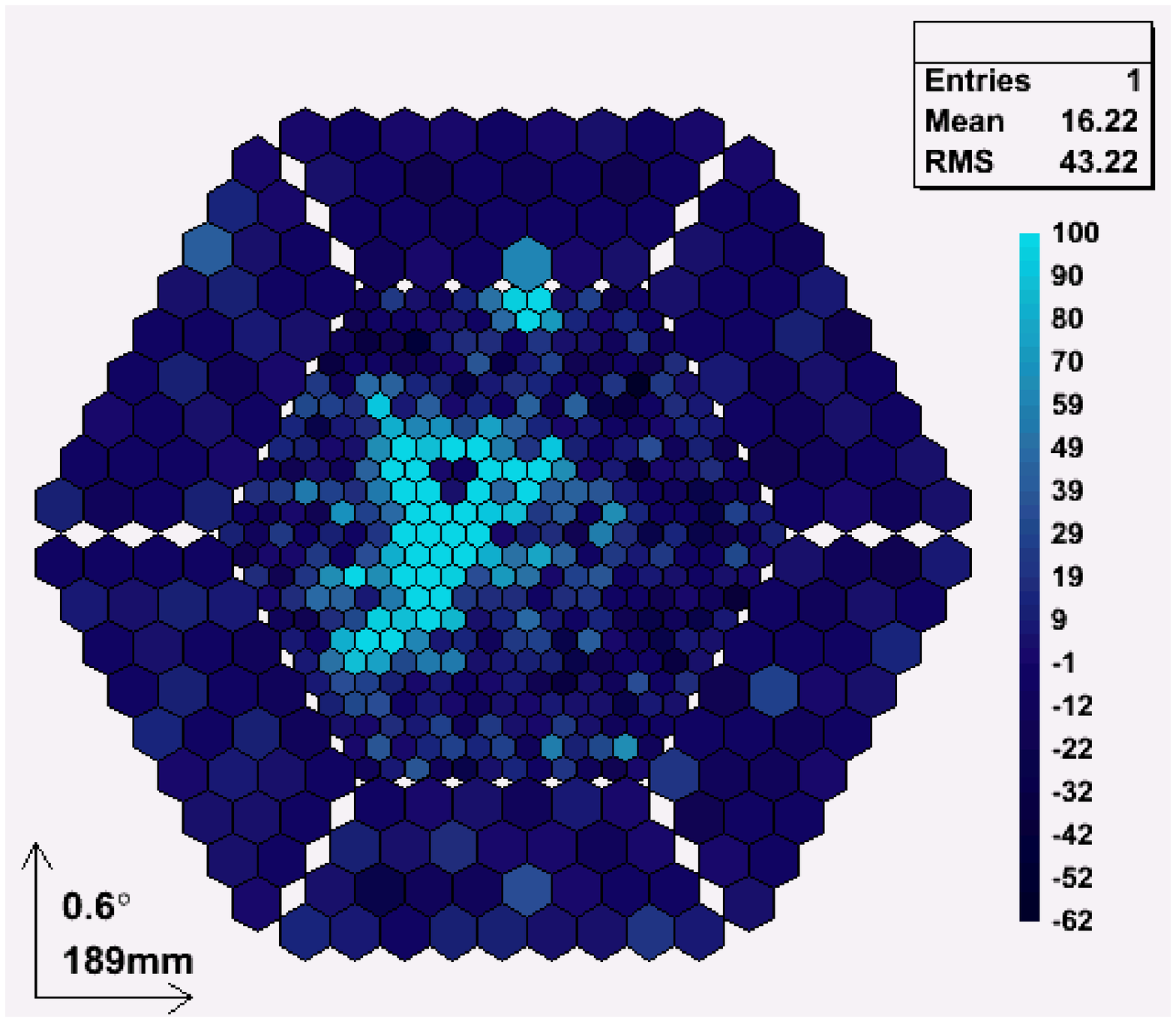}
  \end{center}
  \vspace{-0.5pc}
\end{minipage}
\begin{minipage}[b]{.45\linewidth}
  \begin{center}
    \includegraphics[height=12pc]{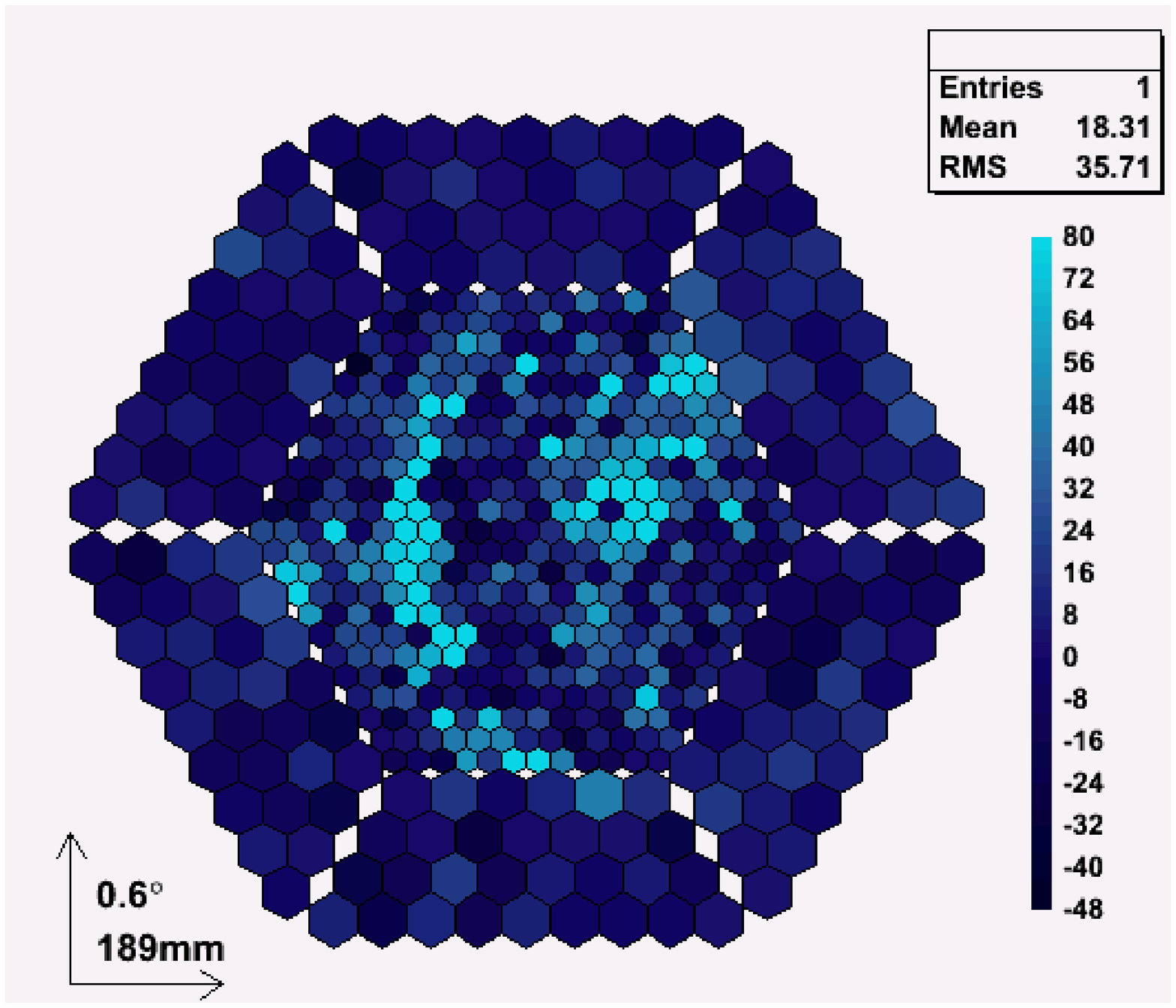}
  \end{center}
  \vspace{-0.5pc}
\end{minipage}\hfill
\begin{minipage}[b]{.45\linewidth}
  \begin{center}
    \includegraphics[height=12pc]{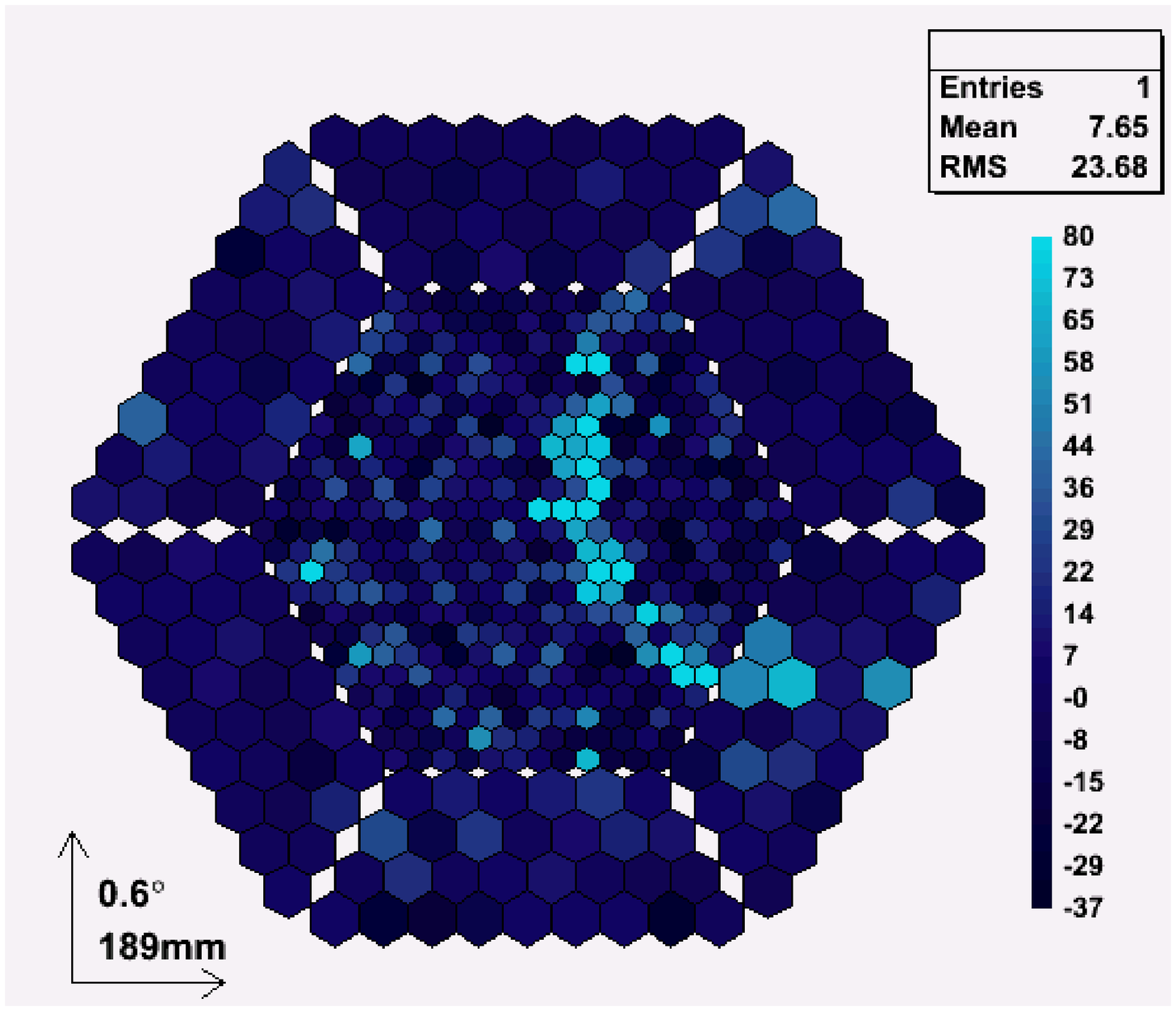}
  \end{center}
  \vspace{-0.5pc}
  \caption{\label{events}
    Displayed are several real events that are good candidates for
  a $\gamma$-ray (top, left), a hadron (top, right), a hadron + muon (bottom,
    left) and an isolated muon (bottom right).}
\end{minipage}

\end{figure}

During the winter we have reached the expected performance of most
of the systems in the telescope. We are presently working on
the evaluation of the global performance of the instrument.

Muons generate very well defined rings in the camera that are used
to calibrate the telescope and to derive the point spread function of the reflector. 
See for instance the event display in the bottom right panel of figure 2.

The point spread function can also be derived by means of the distribution 
of pixel anode currents of bright stars.
In addition the distribution of currents allows us to
monitor the intensity of the NSB in the field of view. NSB generates 
about 1~photoelectron RMS per 3.3~ns FADC sample in the inner PMT pixels.
We also make use of anode currents to estimate the mispointing of the
telescope and to crosscheck the absolute position of the source.

The telescope trigger rate has been found to vary smoothly with
the discrimininator threshold level as expected if it is dominated
by cosmic rays and not by night sky background noise. This happens
for all multiplicities of the trigger (3, 4 and 5~next-neighbours pixels). 
Most of our data have been recorded with a threshold level around 4~phe
and 4~NN trigger multiplicity.

The arrival time of the Cherenkov pulses to the different pixels in the
camera can be determined with a precision better than 1~ns. This allows
to characterize the time profile of the shower and gives another handle
to estimate the direction of the incident $\gamma$-ray.

The dead time of the system has been calculated from the time
difference of consecutive events and is well within the specifications of
the readout system (less than 0.1\%).

\subsection{Data Sample and Analysis}

It is worth to note that the telescope is still in its commissioning phase. 
All the results presented here are preliminary since the configuration 
and performance of the telescope are not the nominal ones yet and 
the analysis software is still in its development phase.

In the months ellapsed since the inauguration of the telescope,
we have mainly concentrated on low zenith angle ($<$40$^{\circ}$) 
observations of standard TeV candles
like Crab Nebula, Mrk~421, Mrk~501, 1ES~1426 and 1ES~1959.
A roughly equivalent amount of OFF source data have been recorded under
the same conditions of the ON data for background substraction.

We have selected only very short data samples for which the weather conditions 
were excellent on the basis of the trigger rates and star extinction 
measurements, all the telescope hardware systems were performing 
nominally and the data have been already preprocessed and calibrated.

The calibration of the telescope has been performed mainly using the
so-called F-Factor method: the number of photoelectrons produced
by the light pulser in a pixel are estimated using the width of the 
charge distribution recorded by the FADCs. 
Images are cleaned using two levels (tail and boundary cuts).

We have applied a standard analysis based on Hillas parameters.
Cuts on WIDTH, LENGTH and DIST have been optimized on a small sample of the
data and then applied to the whole sample. The cuts are dynamical, that is,
they scale with the size of the shower. Any mispointing of the telescope
has been corrected using reference stars in the anode currents and 
the ``false source method'' where the standard Hillas analysis
is performed in a grid of positions in the camera field of view.

A lower cut on the shower size of 2000 photons has been applied. This cut 
selects only high energy showers for which a standard analysis 
based on the Hillas parameters is straightforward.
However this increases substantially the analysis threshold as respect to the 
trigger threshold.

\subsubsection{Crab Nebula}

\begin{figure}
  \caption{\label{berlincrabalpha}
    First detection of the Crab Nebula using the MAGIC telescope.}
    \includegraphics[height=13.pc]{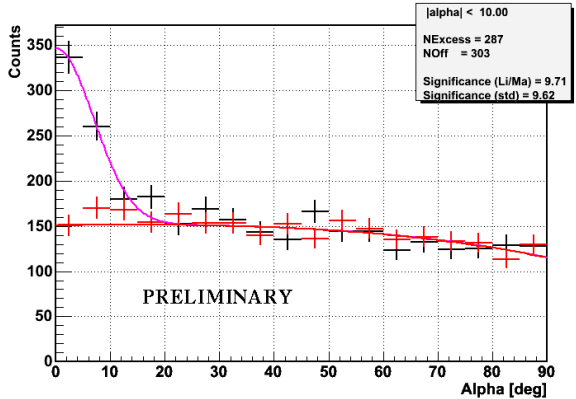}
\end{figure}

The Crab Nebula is a steady emitter at GeV and TeV energies. The $\gamma$-ray
emission is produced by Inverse Compton (IC) scattering of a population of electrons
that are accelerated in the plerion around the central pulsar. The spectrum
of this source has been measured in the GeV range by EGRET and at energies
above 300~GeV by a number of Cherenkov telescopes. The expected level of
emission between 30 and 300 GeV makes it into an excellent calibration candle
for MAGIC. The spectrum in this energy range is interesting by itself
since it allows to further constrain the IC emission parameters (magnetic
field and/or size of the emission region).

Crab was observed during the winter months for about 20 hours under very 
different weather and instrumental conditions.
We have selected a sample of Crab Nebula data of 60 minutes livetime 
at low zenith angle. Fig. 3 displays the distribution
of the ALPHA Hillas parameter for the ON and the corresponding OFF data. 
An excess is clearly visible at low ALPHAs.
We detect 290 gamma candidate events over an expected background
of 300 events. This corresponds to a significance of $\sim$10$\sigma$.

\subsubsection{Mrk~421}

Mrk~421 is one of the so-called TeV BL~Lacs that have been detected above 500~GeV. 
It undergoes the fastest flares that have been observed at these energies,
with flux doubling times as short as 20~minutes. 
The very high energy emission is attributed to IC scattering of electrons
that are accelerated in the base of the jet. The IC peak of the spectral
energy distribution is probably around $\sim$100~GeV. During 2004 the source
has undergone an episode of intense flaring in X-rays. Since the correlation
between the X-ray and the high energy emission is well established, this
source was also an excellent candidate for MAGIC.

\begin{figure}
\begin{minipage}[b]{.50\linewidth}
  \begin{center}
    \includegraphics[height=10.5pc]{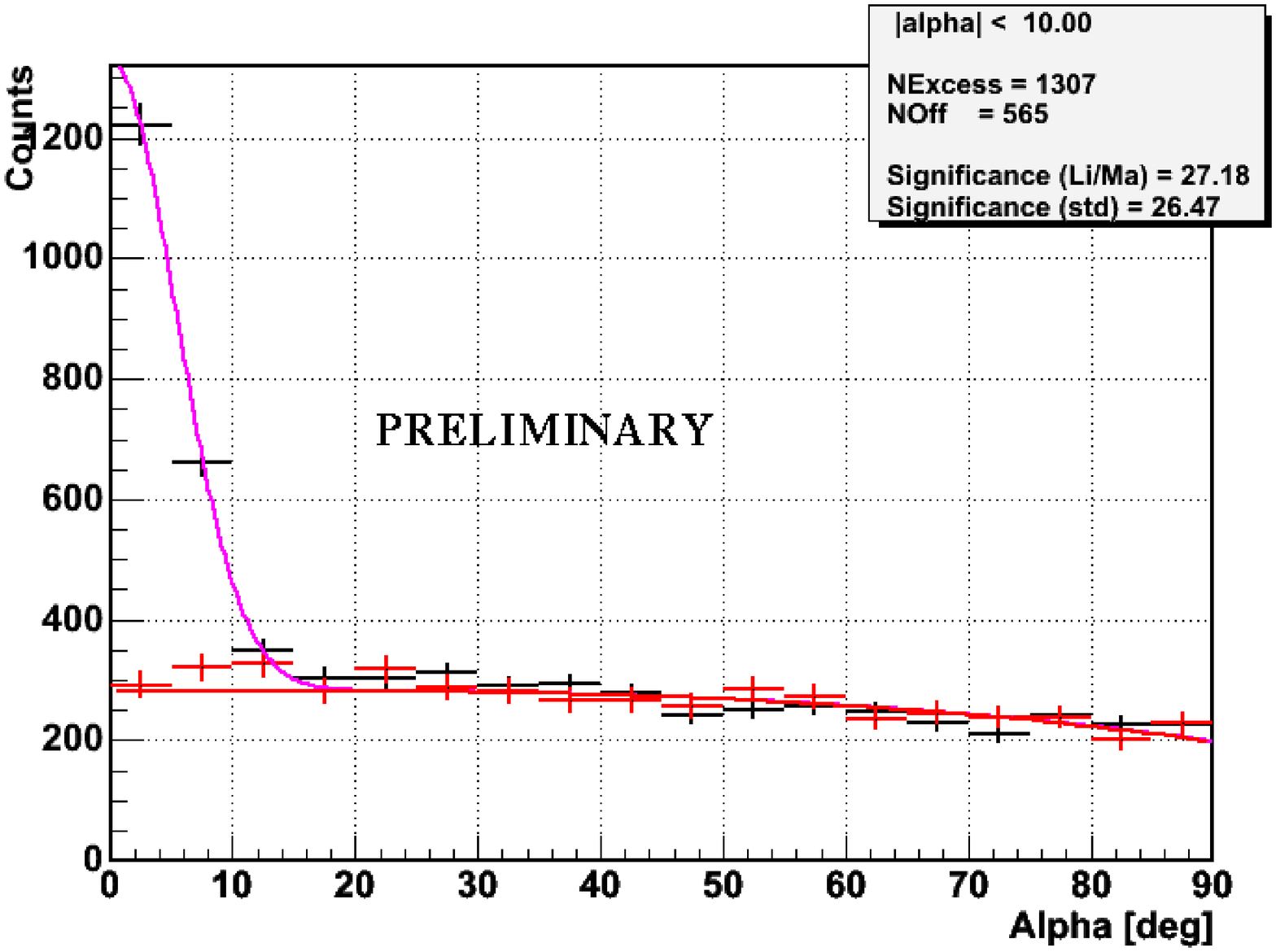}
  \end{center}
  \vspace{-0.5pc}
\end{minipage}\hfill
\begin{minipage}[b]{.50\linewidth}
  \begin{center}
    \includegraphics[height=11.5pc]{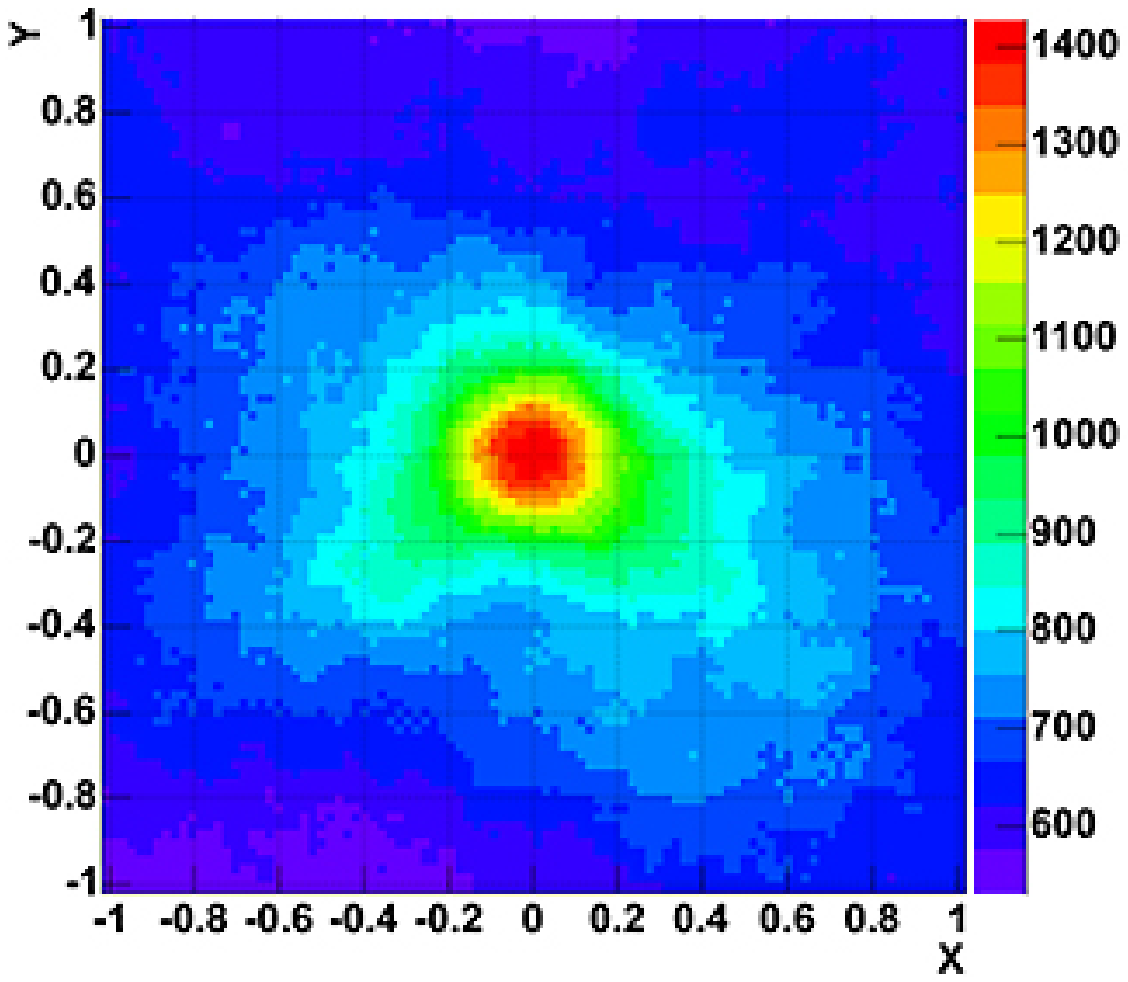}
  \end{center}
  \vspace{-0.5pc}
  \caption{\label{berlin421alpha}
    Left: Alpha plot for the selected sample
    of Mrk~421 data (2004/03/15). Right:
  Result of the false source analysis of the Mrk~421 data sample.}
\end{minipage}
\end{figure}

The 2004 observation campaign started on January and finished on May.
A simple online analysis running on site has shown clear evidence
for signals during most of the campaign. MAGIC also joined
a multiwavelength campaign with VERITAS and RXTE in May 9th. The data
are still not available for full analysis but again showed 
a significant signal in the online analysis.

Here we present the results of a small sample of 96 minutes in optimal
conditions in the night of March 14/15 2004.
Fig. 4 (left) shows the pointing angle ``alpha'' distribution
for these data and the corresponding OFF source data. 
A total of 1307 excess events are observed over an expected background
of 565 events. The significance of the detection is around 27$\sigma$.
It is worth to mention that there was a visible excess in the alpha
distribution even before any cuts.

The false method plot on the right of Fig. \ref{berlin421alpha} 
confirms the detection and the position of the excess
coincides with the source position after pointing corrections.
It must be noted that this plot provides only a rather poor estimate of the
angular resolution of the telescope that is normally calculated using
more sophisticated methods.

\subsubsection{Further Progress}

An estimate of the energy threshold of the telescope during this observation
of the source's $\gamma$-ray flux is under way. We are working to 
improve the significance and the excess of the detection by reducing
the analysis threshold energy and by applying optimized cuts on 
other parameters and using the Cherenkov pulse time profile. These data will be 
complemented soon with the analysis fo the full sample that has been
recorded during the last months.

\section{Future Instrumental Developments}

MAGIC as it stands now has always been thought as the first of a series
of instruments aimed at reducing the energy threshold and increasing 
the sensitivity in the GeV energy range. 

The second instrument in this series is already under construction and 
will have a reflector of the same size of MAGIC, but will incorporate a number 
of new techniques that will improve its performance \cite{Mirzoyan}. Among
them is a faster digitizing system (with $>$2~GHz sampling rate) is in its latest
stage of development. We are also working on several light detectors
with increased efficiency (HPDs and arrays of Geiger mode SiPM) 
that may be incorporated in the camera of the
second telescope or even in the original MAGIC telescope. The frame 
for the second telescope is already under construction and will be delivered
to La Palma in 2005.

The third instrument is ECO-1000 \cite{Merck, Baixeras}, a Cherenkov
telescope with 1000~m$^2$ reflector surface (roughly four times larger than
MAGIC). ECO-1000 intends to take the Cherenkov Telescope technique 
to its extreme and reduce the threshold down to 5~GeV with a
sensitivity of $\sim$3$\cdot$10$^{-10}$cm$^{-2}$s$^{-1}$. This is 
an energy range where it fully overlaps detectors
on board satellites like GLAST. Due to its large collection area and 
high flux sensitivity, ECO-1000 is the ideal instrument to complement GLAST's
huge field of view in the study of transient sources.

\section{Conclusions}

MAGIC, the very large new-generation IACT designed specifically
for the exploration of the 10-300 GeV gamma energy gap is in its
final commissioning phase and is expected to enter regular operation
after the summer.

So far all the new technical components are working 
with their nominal performance, and are now undergoing
extensive checks. The telescope has already detected two
very high energy sources, Crab and Mrk~421, with high significance.

We can conclude that MAGIC is well on its way and entering the
discovery phase. The next months will show the full potential 
of the lowest energy threshold Cherenkov telescope ever built.

\acknowledgements

We would like to thank the IAC for excellent working conditions.
The support of the Italian INFN, German BMBF and Spanish CICYT
is gratefully acknowledged.


\theendnotes

\end{article}
\end{document}